\documentclass[10pt,aps,prl,twocolumn,showpacs,showkeys]{revtex4-1}
\usepackage{amsmath}
\usepackage[abs]{overpic}
\usepackage{graphicx}
\usepackage{dcolumn}
\usepackage{bm}
\usepackage{color}
\usepackage{hyperref}
\hypersetup{backref=true,pagebackref=true,hyperindex=true,colorlinks=true,breaklinks=true,citecolor=blue,urlcolor=blue,linkcolor=blue,bookmarks=true,bookmarksopen=true}
\begin{document}

\title{Phenomenology of chiral damping in noncentrosymmetric magnets}
\date{\today}
\author{Collins Ashu Akosa$^1$}
\author{Ioan Mihai Miron$^{2,3,4}$}
\author{Gilles Gaudin$^{2,3,4}$}
\author{Aur\'elien Manchon$^1$}
\email{aurelien.manchon@kaust.edu.sa}
\affiliation{$^1$King Abdullah University of Science and Technology, Physical Science and Engineering Division, Thuwal 23955, Saudi Arabia}
\affiliation{$^2$Univ. Grenoble Alpes INAC-SPINTEC F-38000 Grenoble France}
\affiliation{$^3$CNRS INAC-SPINTEC F-38000 Grenoble France}
\affiliation{$^4$CEA INAC-SPINTEC F-38000 Grenoble France}

\begin{abstract}
A phenomenology of magnetic chiral damping is proposed in the context of magnetic materials lacking inversion symmetry breaking. We show that the magnetic damping tensor adopts a general form that accounts for a component linear in magnetization gradient in the form of Lifshitz invariants. We propose different microscopic mechanisms that can produce such a damping in ferromagnetic metals, among which spin pumping in the presence of anomalous Hall effect and an effective "$s$-$d$" Dzyaloshinskii-Moriya antisymmetric exchange. The implication of this chiral damping in terms of domain wall motion is investigated in the flow and creep regimes. These predictions have major importance in the context of field- and current-driven texture motion in noncentrosymmetric (ferro-, ferri-, antiferro-)magnets, not limited to metals. 
\end{abstract}

\maketitle

Understanding energy relaxation processes of fast dissipative systems at the nanoscale is of paramount importance for the smart design and operation of ultrafast nano devices. In this respect, magnetic heterostructures have drawn increasing enthusiasm in the past ten years with the optical control of magnetic order at the sub-picosecond scale \cite{kimel2004,fullerton} and the promises of ultrafast domain wall motion in asymmetrically grown metallic multilayers \cite{miron2011a}. In fact, the observation of ultrahigh current-driven velocity in ultrathin multilayers with potentially strong disorder has come as a surprise and triggered intense investigations on the physics emerging from symmetry breaking in strong spin-orbit coupled magnets  \cite{emori,shyang}. These studies have unravelled the essential role played by Dzyaloshinskii-Moriya interaction \cite{d,m} (DMI), an antisymmetric exchange interaction that emerges in magnets lacking spatial inversion symmetry. This interaction forces neighboring spin to align perpendicular to each other and competes with the ferromagnetic exchange resulting in distorted textures such as spin spirals or skyrmions, as observed in bulk inversion asymmetric magnets \cite{nagaosa} (B20, ZnS or pyrochlores three-dimensional crystals) as well as at the interface of transition metals \cite{bode2007,heinze2011} (Mn/W, Fe/W, Pt/Co etc.). In perpendicularly magnetized domain walls, this interaction favors N\'eel over Bloch configuration \cite{heide,chen,tetienne}, a key ingredient explaining most of the thought-provoking observations reported to date \cite{thiaville}. The dynamics of spin waves can also be modified by DMI, which distorts the energy dispersion \cite{zakeri2010} and results in a relaxation that depends on the propagation direction \cite{zakeri2012,moon2013}. 

Another crucial aspect of fast dynamical processes is the nature of the energy relaxation. In the hydrodynamic limit of magnetic systems, this dissipation is written in the form of a non-local tensor [see Eq. (\ref{eq:llg})] whose complex physics is associated with a wide variety of mechanisms such as many-magnon scattering \cite {mills} and itinerant electron spin relaxation \cite{kambersky}. Since the energy relaxation rate of spin waves depends on their wave vector (the higher the spin wave energy, the stronger its dissipation), the magnetic damping of smooth magnetic textures (i.e. in the long wavelength limit $q$) depends on the inverse of the exchange length, $q\sim 1/\Delta$. In inversion symmetric systems, this results in a correction to the magnetic damping of the order of $1/\Delta^2$ \cite{zhangzhang,nembach,Yuan}. However, in magnetic systems lacking inversion symmetry such as the systems in which DMI is observed (i.e. B20 and ZnS crystals and transition metal interfaces), one can reasonably expect that the energy dissipation becomes chiral, namely that a component linear in the magnetization gradient emerges thereby fulfilling Neumann's principle stating that "any physical properties of a system possesses the symmetry of that system".

In the present work, we phenomenologically explore the nature of the magnetic damping in noncentrosymmetric magnets and reveal that spatial inversion symmetry breaking results in the emergence of such a chiral damping that vanishes when the symmetry of inversion is recovered. The inclusion of such a chiral damping in the equation of motion of magnetic textures opens appealing avenues to solve recent puzzling observations that can not be fully accounted for with DMI only.\par


{\em Symmetry considerations} - The equation of motion governing the dynamics of continuous magnetic textures is given by the extended Landau-Lifschitz-Gilbert (LLG) equation
\begin{equation}\label{eq:llg}
\partial_t {\bf m} = - \gamma {\bf m}\times{\bf H}_{\rm eff} + {\bf m}({\bf r})\times \int d{\bf r'}\underline{\underline{\boldsymbol{\alpha}}}({\bf r},{\bf r}')\partial_t{\bf m}({\bf r'}),   
\end{equation}
where ${\bf m}({\bf r},t) = {\bf M}({\bf r},t)/M_s$ is a unit vector in the direction of the magnetization ${\bf M}({\bf r},t)$, $\gamma$ is the gyromagnetic ratio, ${\bf H}_{\rm eff}$ is the effective field incorporating the external applied, anisotropy, exchange, DMI and demagnetizing fields, and $\underline{\underline{\boldsymbol{\alpha}}}({\bf r},{\bf r}')$ is the magnetic damping expressed as a non-local second-rank tensor. In the limit of smooth textures, the tensorial components of the damping is a function of the magnetization direction and of its spatial gradients, $\alpha^{ij} = \alpha^{ij}({\bf m}, \nabla{\bf m})$. Performing an expansion up to the first order in magnetization gradient, one obtains (see also \cite{Kjetil})
\begin{equation}\label{eq:chir}
\alpha^{ij}   = \alpha^{ij}_0 + \sum_{lm} K_{lm}^{ij}m_lm_m + \sum_{klm} L_{klm}^{ij}m_k \partial_l m_m.
\end{equation}
The first term is the isotropic damping, the second term ($\sim K_{lm}^{ij}$) amounts for the anisotropy arising from the crystalline environment and the third term is the chiral damping. It should be noted that only terms bilinear in magnetization direction $m_i$, i.e. even under time reversal symmetry, are retained in the expansion. Since the focus of this study is on the chiral nature of magnetic dissipation, we ignore the anisotropy term $( \propto K_{lm}^{ij} )$ at this stage. Spatial inversion symmetry breaking imposes the third term of Eq. (\ref{eq:chir}) to reduce to a sum of Lifshitz antisymmetric invariants, $\propto m_k\partial_lm_m - m_m\partial_l m_k $ (i.e. $L_{klm}^{ij} = -L_{mlk}^{ij}$). To illustrate the general form of this chiral damping, we consider two prototypical systems. In the case of a \textit{cubic} three-dimensional system with bulk spatial inversion symmetry breaking (e.g. B20 or ZnS crystals), all the three directions (${\bf x}$ , ${\bf y}$ and ${\bf z}$) are equivalent, and the chiral damping adopts the general form 
\begin{equation}
\alpha^{ij} = \alpha^{ij}_0 + \alpha^{ij}_{3d}\Delta {\bf m}\cdot[\boldsymbol{\nabla}\times {\bf m}],
\end{equation}
where $\Delta$ defines the characteristic exchange length. In a two-dimensional system with interfacial symmetry breaking along ${\bf z}$, i.e. invariant under $\mathcal{C}_{\infty z}$ rotation symmetry, the damping takes the form 
\begin{equation}\label{eq:cd}
\alpha^{ij} = \alpha^{ij}_0 + \alpha^{ij}_{z}\Delta {\bf m}\cdot[({\bf z}\times\boldsymbol{\nabla})\times {\bf m}].
\end{equation}

As dictated by Neumann's principle, the chiral damping possesses the same symmetry as DMI given by Nagaosa {et. al.} \cite{nagaosa} and Thiaville {et. al.} \cite{thiaville}, respectively. In addition, Onsager reciprocity imposes that $\alpha^{ij}=\alpha^{ji}$. Therefore, using simple symmetry arguments, one can construct a chiral damping up to linear order in magnetization gradient. These considerations suggest that such a damping is present in noncentrosymmetric (ferro-, ferri- and antiferro-)magnets in general, not limited to metals. However, this phenomenology does not provide information regarding the strength of the chiral damping itself, the relative values of the off-diagonal tensor elements ($\sim\alpha^{i\neq j}$) compared to the diagonal ones  ($\sim\alpha^{ii}$) nor does it indicate the underlying physical mechanisms responsible for it. Let us now turn our attention towards the microscopic origin of such chiral damping.\par


{\em Microscopic origin of the chiral damping} - In this work, we focus our attention on magnetic textures in metals where the magnetic damping is driven by the spin relaxation of itinerant electrons \cite{kambersky}. In ferromagnets with interfacial Rashba spin-orbit coupling, the magnetic damping adopts the form of a tensor linear in both magnetization gradient and Rashba strength, as derived by Kim et al. \cite{kim2012} and Wang et al. \cite{wang2014}. In this model, the magnetic dissipation is mediated by the same spin-orbit coupled itinerant electrons which mediate the DMI on the local magnetic moment (see e.g. Ref \cite{kim2013}). Besides this effect, we here propose two additional mechanisms that can contribute to the chiral damping.\par

The first mechanism arises from the interplay between spin motive force and anomalous Hall effect. It has been recently shown that time-dependent spin textures generate a local spin current \cite{barnes2007}, ${\bf J}_i^s = \frac{g\mu_B\hbar G_0}{4e^2}\partial_t{\bf m}\times\partial_i{\bf m}$, flowing along the direction of the texture ${\bf e}_i$, polarized along $\partial_t{\bf m}\times\partial_i{\bf m}$ ($G_0$ is the electrical conductivity), and that induces a magnetic damping at the second order of spatial gradient \cite{zhangzhang}. When anomalous Hall effect is present in the ferromagnet, this primary spin current ${\bf J}_i^s$ is converted into a secondary spin current  ${\bf J}_j^{s} = \theta_{\rm H} P [{\bf J}_i^{s}\cdot({\bf e}_i\times{\bf e}_j)]{\bf e}_i\times{\bf e}_j$, flowing along ${\bf e}_j$ and polarized along ${\bf e}_i\times{\bf e}_j$. Here, $\theta_{\rm H}$ is the spin Hall angle and $P$ is the spin polarization in the ferromagnet. This secondary spin current can be injected in an adjacent spin sink with strong spin relaxation, thereby inducing a damping torque on the magnetization, similar to the spin pumping mechanism \cite{Tserkovnyak}. Considering a one-dimensional domain wall along ${\bf x}$ deposited on a heavy metal with an interface normal to ${\bf z}$, we obtain a damping torque on the form \cite{supp}
\begin{equation}\label{eq:ahe}
{\bm \tau}=\theta_{\rm H}{\cal A}\frac{g\mu_B\hbar PG_0}{4e^2}[(\partial_t{\bf m}\times\partial_i{\bf m})\cdot{\bf y}]{\bf m}\times({\bf y}\times{\bf m}),
\end{equation}
${\cal A}$ being a renormalization factor arising from the spin current backflow. This damping torque is proportional to $\sim\sin2\varphi$, $\varphi$ being the azimuthal angle of the magnetization, and vanishes when the wall is either in Bloch ($\varphi=0$) or N\'eel configuration ($\varphi=\pi/2$).\par

The second mechanism is directly related to DMI. In transition metal ferromagnets, the orbital characters of the delocalized ($spd$ hybridized) and localized electrons ($pd$ hybridized) are mixed so that both types of electrons contribute to the magnetic exchange. This is particularly true in the case of DMI: {\em ab initio} calculations indicate that interactions beyond the next-nearest neighbor contribute significantly to the total DMI \cite{kashid}, proving that delocalized orbitals are crucial in determining the overall strength of DMI. Therefore, by parsing the total spin ${\bf S}_i$ into localized ($d$-dominated) and delocalized ($s$-dominated) contributions, ${\bf S}_i = {\bf S}_i^{ d} + \hat{{\bf s}}_i^{ s}$, the DMI between sites $i$ and $j$ can be phenomenologically rewritten ${\bf D}_{ij} \cdot {\bf S}_i\times{\bf S}_j={\bf D}_{ij}^{ dd} \cdot {\bf S}_i^{ d}\times\hat{{\bf S}}_j^{ d} +{\bf D}_{ij}^{ sd} \cdot {\bf S}_i^{ d}\times\hat{{\bf s}}_j^{ s}$. The first term only involves orbital overlap between localized states while the second term describes the chiral exchange between the local spin and the itinerant spin. In the continuous limit, the Hamiltonian of the itinerant electron can be then written

\begin{eqnarray}\label{eq:ham3}
\hat{\mathcal{H}}_{\rm sd} &=&  \frac{\hat{\bf p}^2}{2m}+J_{\rm ex} {\bf m}\cdot \hat{\boldsymbol{\sigma}}\nonumber \\&&+\frac{D}{\hbar} [({\bf z}\times\hat{\bf p}) \times {\bf m}]\cdot \hat{\boldsymbol{\sigma}} +\frac{\alpha_{\rm R}}{\hbar} ({\bf z}\times\hat{\bf p})\cdot \hat{\boldsymbol{\sigma}},
\end{eqnarray}
where $\hat{\boldsymbol{\sigma}}$ is the vector of Pauli spin matrices, $J_{\rm ex}$, $D$ and $\alpha_{\rm R}$ are the strength of the exchange, $s$-$d$ DM and Rashba spin-orbit interactions respectively. In other words, because of the magnetic texture the itinerant electron spin experiences an additional effective field of the form $\sim ({\bf z}\times\hat{\bf p}) \times {\bf m}$. This introduces a chirality in the magnetic damping mediated by itinerant electrons.\par 
 
From this point, deriving the effective magnetic damping follows the standard procedure. One can extract the equation of motion of the itinerant spins from Eq. (\ref{eq:ham3}) and define the spin current induced by the moving magnetic texture (see Supplementary Materials \cite{supp}). Following this method, we obtain a semi-classical Bloch equation for the itinerant electron spin density ${\bf s} = \langle\hat{\boldsymbol{\sigma}}\rangle$  as
\begin{equation}\label{eq:scon}
\partial_t{\bf s} + \boldsymbol{\nabla}\cdot \mathcal{J} = -\frac{_1}{{^{\tau_{\rm ex}}}} {\bf s}\times{\bf m} - \frac{_\Delta}{^{{\tau_{\rm D}}}} ({\nabla}_{\bf z}\times{\bf m})\times{\bf s} - \frac{_\Delta}{{^{\tau_{\rm R}}}} {\nabla}_{\bf z}\times{\bf s} - \Gamma_{\rm re},
\end{equation}
where ${\bm\nabla}_{\bf z}={\bf z}\times{\bm\nabla}$, $\Gamma_{\rm re}$ represents the spin relaxation and dephasing, $\Delta$ is the exchange length, $\tau_{\rm ex}=\hbar/2J_{\rm ex}$ is the spin precession time,  $\tau_{\rm D}=\hbar\Delta/D$ and $\tau_{\rm R}=\hbar\Delta/\alpha_{\rm R}$ are the characteristic time scales for the DMI and Rashba interaction respectively. $\mathcal{J} = -\mathcal{D} \boldsymbol{\nabla}\otimes {\bf s}$ is the spin current density tensor, $\mathcal{D}$ being the diffusion constant. Let us now write the spin density in the form ${\bf s} = n_s{\bf m} + \delta {\bf s}$, where $n_s$ ($\delta{\bf s}$) is the (non-)equilibrium spin density, and assume the relaxation time approximation such that $\Gamma_{\rm re}({\bf s}) = \frac{1}{\tau_{\rm sf}} \delta{\bf s} + \frac{1}{\tau_{\rm \varphi}} {\bf m}\times(\delta{\bf s}\times{\bf m})$, accounting for the spin-flip relaxation ($\sim \tau_{\rm sf}$) and the spin dephasing  ($\sim \tau_{\rm \varphi}$). After some algebra \cite{supp}, one obtains the torque $\bm \tau$ generated by a precessing magnetization $\sim \partial_t{\bf m}$
  \begin{eqnarray}\label{eq:tdn2}
\boldsymbol{\tau}/\tilde{n}_s &\approx& (1+\chi\xi-\beta{\bf m}\times)\left[ -\partial_t{\bf m}\right.\\\nonumber
&&+\lambda_{\rm D}[(({\bf z}\times{\bm\nabla})\times{\bf m})\times({\bf m}\times\partial_t{\bf m}+\xi\partial_t{\bf m})]_\bot\\
&&\left.+\lambda_{\rm R}[({\bf z}\times{\bm\nabla})\times({\bf m}\times\partial_t{\bf m}+\xi\partial_t{\bf m})]_\bot\right].\nonumber
\end{eqnarray}
In this expression, $\beta=\tau_{\rm ex}/\tau_{\rm sf}$, $\chi=\tau_{\rm ex}/\tau_{\varphi}$, $\xi=\chi+\beta$, and $\tilde{n}_s=n_s/(1+\xi^2)$ and the subscript $\bot$ indicates that the torque is defined perpendicular to the magnetization ${\bf m}$. The first term has been derived previously \cite{zhangli}, the second term ($\lambda_{\rm D}= \Delta\tau_{\rm ex}/\tau_{\rm D}$) arises from the $s$-$d$ DMI exchange, and the third term ($\lambda_{\rm R}=\tau_{\rm ex}/\tau_{\rm R}$) arises from Rashba spin-orbit coupling \cite{kim2012,wang2014}. The total torque $\bm \tau$ contributes both to the renormalization of the gyromatic ratio (terms that preserve time-reversal symmetry) and to dissipation (terms that break time-reversal symmetry). 

{\em Domain wall motion} - To illustrate the effect of this chiral damping on the dynamics of magnetic textures in noncentrosymmetric metals, we first derive the equation of motion of a one-dimensional perpendicular domain wall, such as the ones commonly observed in heavy metal/ferromagnet asymmetric multilayers \cite{miron2011a,emori,shyang}. The magnetization is defined ${\bf m}=(\cos\varphi\sin\theta,\sin\varphi\sin\theta,\cos\theta)$, where $\varphi = \varphi(t)$ is the azimuthal angle and $\theta(x) = 2\tan^{-1}\left(\exp[s(x-X)/\Delta]\right)$, $X$ being the domain wall centre, and $s=\pm 1$ defining the domain wall chirality ($\uparrow \downarrow$ or $\downarrow\uparrow $, respectively). We consider a magnetic domain wall submitted to a magnetic anisotropy field ${\bf H}_{k}=H_{k}\sin\theta\sin\varphi{\bf y}$, favoring Bloch configuration and an applied magnetic field ${\bf H}=H_x{\bf x}$ favoring N\'eel configuration. The damping torque is given by Eq. (\ref{eq:tdn2})(for the sake of simplicity, we do not consider the influence of the torque arising from anomalous Hall effect, Eq. (\ref{eq:ahe}), in this work). The rigid domain wall dynamics is described by the coupled equations

\begin{eqnarray}  \label{eq:vv}
\frac{s\partial_\tau x}{\Delta} &=&   \frac{\pi}{2}(-H_x\sin\varphi+\frac{H_k}{2}\sin2\varphi)+(\alpha-s\nu\cos\varphi) { H}_{\rm z} ,\nonumber\\
\partial_\tau \varphi  &=&-s\frac{\pi}{2}(\alpha-s\mu\cos\varphi)(-H_x\sin\varphi+\frac{H_k}{2}\sin2\varphi)+{ H}_{\rm z},\nonumber\\
&&
\end{eqnarray}
where we defined $\tau=\gamma t$, $\mu=(\lambda_{\rm D}-\xi\lambda_{\rm S})(\pi\tilde{n}_s/4\Delta)$, and $\nu=(\beta\lambda_{\rm D}+\lambda_{\rm S})(\pi\tilde{n}_s/4\Delta)$ ($\alpha,\nu,\mu\ll 1$). In Eq. (\ref{eq:vv}), we neglected the components of the torque ${\bm\tau}$ that are even in magnetization (i.e. that renormalizes the gyromagnetic ratio) and only consider the dissipative components. The damping due to $s$-$d$ DMI and Rashba SOC both produce a contribution proportional to $s\cos\varphi$, i.e. it depends on the domain wall chirality $s$ as well as on the direction of the azimuthal angle $\varphi$. Notice that the DM field does not explicitly enters these equations since it can be simply modeled by a chiral in-plane field $\sim sH_x$ \cite{thiaville}.

\begin{figure}[t]
\centering
\includegraphics[width=9cm]{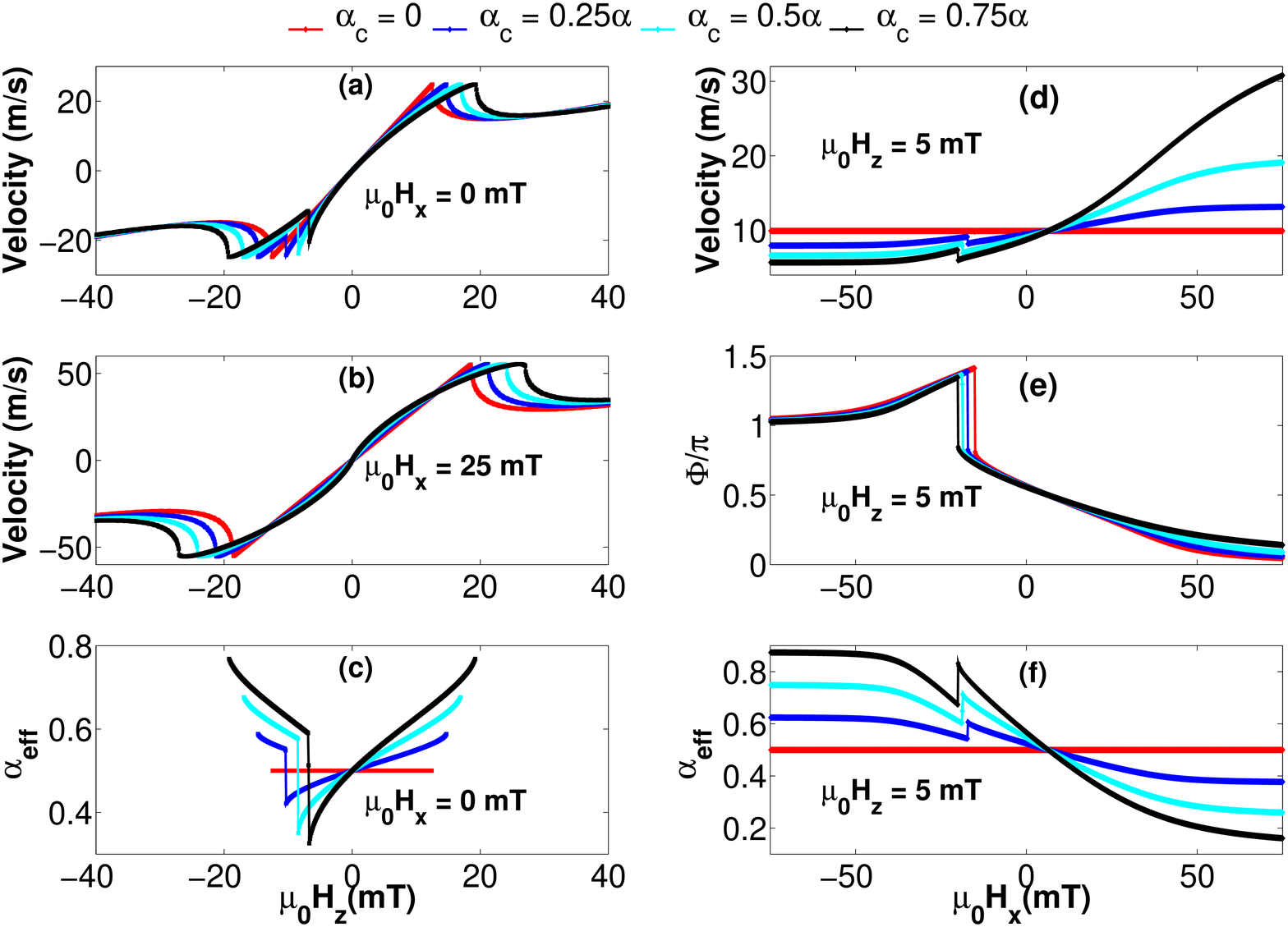}
\caption{(Color online) (a,b) Domain wall velocity as a function of perpendicular magnetic field for different chiral damping strengths, at $H_x=$0 mT and $H_x=$25 mT, respectively; (c) corresponding effective damping. (d) Domain wall velocity as a function of in-plane magnetic field for different chiral damping strengths, and for $H_z=$5 mT; (e,f) corresponding azimuthal angle and effective damping, respectively. The parameters are $\alpha$= 0.5 , $H_{\rm WB}$ = 12.5 mT.
 \label{fig:fig1}}
\end{figure}

Let us now investigate the influence of this damping on the field-driven motion of a domain wall submitted to both perpendicular ($H_z$) and in-plane ($H_x$) magnetic fields. In this example, we chose $\mu=\nu=\alpha_{\rm c}$ for simplicity. Figure \ref{fig:fig1}(a) shows the (time-averaged) velocity of the domain wall as a function of $H_z$ for difference chiral damping strengths and $H_x=0$. These velocity curves display the usual Walker breakdown around $H_z\approx 12-19$ mT, with an additional kink at negative $H_z$ directly attributed to the chiral damping. Below Walker breakdown and in the absence of in-plane field $H_x$, the domain wall azimuthal angle $\varphi$ obeys $H_z = s\frac{\pi}{4}H_k(\alpha - s\alpha_c \cos\varphi)\sin2\varphi$, which produces a kink around $H_z\approx s\alpha_cH_{\rm WB}/\alpha$ where $H_{\rm WB}=\alpha \pi H_k/4$, which is associated with a jump in the effective damping, $\alpha_{\rm eff}=\alpha-\alpha_c\cos\varphi$ [see Fig. \ref{fig:fig1}(c)]. When applying a large in-plane field $H_x$, the azimuthal angle is $\varphi\approx 0$ (N\'eel configuration) and the damping does not depend on $H_z$. Hence, the kink disappears as shown in Fig. \ref{fig:fig1}(b). More interestingly, when the domain wall is tuned from Bloch to N\'eel configuration by an in-plane field [see Figs. \ref{fig:fig1}(d)-(f)], the effective damping is strongly modified [Fig. \ref{fig:fig1}(f)] and domain wall velocity becomes strongly asymmetric as reported on Fig. \ref{fig:fig1}(d). Notice that the kink is still observable at small negative in-plane field.

We now turn our attention towards the creep regime, which is of most importance for field-driven domain wall motion in ultrathin disordered multilayers. In this case, the creep law predicts \cite{chauve2000}
\begin{equation}\label{eq:me}
v_{\rm creep}=v_0\exp\left(-(T_c/T)(H_c/H_z)^{1/4}\right)
\end{equation}
where $T_c$ is the critical temperature (close to the depinning temperature in fact), $T$ is the sample temperature, $H_z$ is the applied field and $H_c$ is the critical field needed to overcome the disorder pinning potential. This expression assumes $H_z\ll H_c$, as well as $T_c\gg T$. The exponent gathers only terms accounting for the disordered energy landscape of the system and does not include any dissipative contributions in principle. In contrast, the coefficient $v_0$ depends on both the energy landscape as well as on the viscosity of the elastic wall. In the limit of an {\em overdamped} membrane (i.e., when $\alpha\partial_t{\bf m}\gg\partial_t{\bf m}$), one can show that $v_0\approx \gamma \Delta H_c/\alpha$ \cite{chauve2000}. It is mostly probable that even for strongly disordered ferromagnets, the creeping domain wall is not in the overdamped regime, although very large damping ($\alpha\approx 0.5$) have been reported in Pt/Co/AlOx and Pt/Co/Pt systems \cite{miron2011a}. However, since the theoretical description of the creep motion of intermediate damped systems is not fully understood, we propose to investigate the impact of chiral damping in this limit.\par

\begin{figure}[ht!]
\centering{
\includegraphics[width=3.0in]{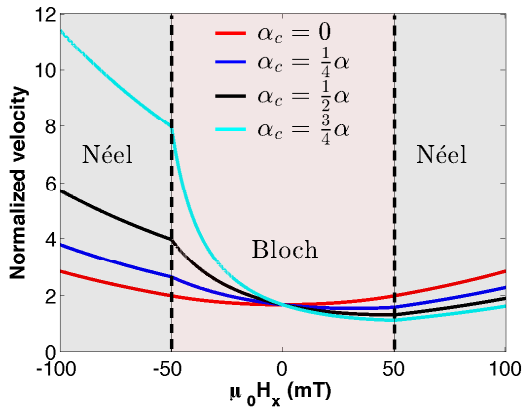}
\caption{(Color online) Normalized domain wall velocity as a function of in-plane magnetic field $H_x$ in the presence of a driving field $H_z$ = 20mT. In-plane magnetic field pushes the domain from a Bloch to a N\'eel configuration and hence increases the azimuthal angle $\varphi$ thereby modifying the damping of the wall. A kink in the velocity is observed at $H_x$=$H_k$, when the domain wall saturates in the N\'eel configuration when the damping becomes independent on the in-plane field $H_x$.} \label{fig:fig3}}
\end{figure}

To evaluate the impact of the chiral damping, we follow the procedure proposed by Je et al. \cite{je} and rewrite the creep law
\begin{equation}
v_{\rm creep}({\bf H}) =\eta_0\frac{\sigma_{\rm DW}(H_x)^\beta}{\alpha_{\rm eff}(H_x)} \exp\left(-\chi_0\frac{\sigma_{\rm DW}^{1/4}(H_x)}{\sigma_{\rm DW}^{1/4}(0)}H_z^{-\frac{1}{4}}\right).
\end{equation}
Here, $\sigma_{\rm DW}$ is the (chiral) energy of the domain wall and $\beta$ is a coefficient that describes the scaling law between the critical force $H_c$ and the domain wall energy and $\eta_0$ and $\chi_0$ are normalization factors which can be chosen to fit experimental data of Ref. \cite{je}. In the phenomenological model adopted here, the domain wall energy reads

\begin{equation}\label{eq:me}
\sigma_{\rm DW} = \sigma_0 + \pi \Delta \mu_0 M_s\left[\frac{_1}{^2}H_k \cos\varphi -  H_x\right]\cos\varphi.
\end{equation}
In the right-hand side of Eq. (\ref{eq:me}), the first term accounts for the $\varphi$-independent contribution to the magnetic energy,  the second term is the in-plane magnetic anisotropy favoring Bloch configuration and the last term is the in-plane longitudinal magnetic field favoring N\'eel configuration. The energy minimization $\partial_\varphi \sigma_{\rm DW} = 0$ gives \cite{je}
\begin{equation}\label{eq:abl}
 \cos\varphi  = \left\{
  \begin{array}{lr}
    H_x /H_k & : |H_x| \le H_k, \\
    {\rm sign}(H_x /H_k) & : |H_x| > H_k. 
  \end{array}
\right.
\end{equation}
and the corresponding domain wall energy

\begin{equation}\label{eq:dwe}
\frac{\sigma_{\rm DW}}{\sigma_0} = \left\{
  \begin{array}{lr}
    1 - (H_x )^2/H_{\rm DW} H_k& : |H_x| \le H_k \\
     1 + (H_k -2|H_x |)/H_{\rm DW}& : |H_x| > H_k 
  \end{array}
\right.
\end{equation}
where $H_{\rm DW} = \frac{\pi}{2} \Delta \mu_0 M_s/\sigma_0$. The magnetic damping $\alpha_{\rm eff}$ is written in the simplest form $\alpha_{\rm eff} = \alpha + s\alpha_c \cos\varphi$. To investigate the impact of chiral damping on the creep motion, we chose $H_{\rm DW} =1T$, and $H_k =50mT$. The normalized velocity $v(H_x)/v(0)$ of a domain wall as a function of the in-plane field and for different strengths of chiral damping $\alpha_c $ is represented on Fig. \ref{fig:fig3}. It shows three distinct regions: a smooth variation of the velocity in the intermediate field region, $|H_x|<H_k$, where the domain wall is changed from one N\'eel chirality to another, as well as two external region, $|H_x|>H_k$, where the domain wall remains in the N\'eel configuration. In this case, the velocity increases following the exponential law given above. Notice that in this regime and for such a one-dimensional domain wall, the DMI results in an effective in-plane magnetic field whose sign depends on the chirality of the wall, i.e. $H_x\rightarrow H_x-s H_{\rm DMI}$ (see e.g. Ref. \cite{je}). Thus, including DMI in the calculation only results in a horizontal shift of the velocity curve in Fig. \ref{fig:fig3}. \par

We acknowledge that the present phenomenology remains essentially qualitative and although the orders of magnitude discussed in this work are globally consistent with the experimental observations, a microscopic theory of chiral damping using, for instance, density function theory techniques, as well as a more comprehensive model of the creep motion of magnetic domain walls are highly desirable to quantitatively confront with experiments. Nonetheless, the symmetry principles discussed in this Letter are quite general and ensure the existence of such a chiral damping in any magnetic structures (ferromagnets, antiferromagnets, chiral magnets, but also metals and insulators etc.) presenting spatial inversion symmetry breaking. The physical mechanisms responsible for this chiral damping can be spin-orbit coupling but also dipolar coupling or magnetic frustrations, as in the case of DMI. Recent indications of such an asymmetric damping of spin waves \cite{zakeri2012}, correlated with a strong magnonic Rashba effect \cite{zakeri2010} are encouraging indications that such an effect is reasonable and call for more systematic investigations of chiral damping in noncentrosymmetric magnets.

C.A. Akosa and A. Manchon acknowledge support from King Abdullah University of Science and Technology (KAUST).

\end{document}